\documentclass[12pt]{article}

\usepackage{amsfonts,amsthm,amssymb}
\usepackage{graphicx,psfrag}
  
  \newcommand{\field}[1]{\mathbb{#1}}
  \newcommand{\rbb}{\field{R}}
  \newcommand{\cbb}{\field{C}}
  \newcommand{\nbb}{\field{N}}
  \newcommand{\zbb}{\field{Z}}
  \newcommand{\one}{\field{I}}
  
  \newcommand{\im}{\mathrm{i}}
  \newcommand{\oh}{{\textstyle\frac{1}{2}}}

  \newcommand{\tr}{\mathrm{Tr}}
  \newcommand{\der}{\mathrm{d}}
  \newcommand{\wed}{\!\wedge\!}
  \newcommand{\al}{\mathcal{A}}
  \newcommand{\End}{\mathrm{End}}
  
  \newcommand{\lie}{\mathfrak{g}}
  \newcommand{\first}{\mathbf{F}}
  \newcommand{\poly}[1]{\mathcal{H}_{#1}}
  \newcommand{\Adj}{\mathrm{Ad}}
  
  \newcommand{\Vef}{\mathrm{Der}}
  
  \newcommand{\beq}{\begin{equation}}
  \newcommand{\eeq}{\end{equation}}
  
  \newtheorem{theorem}{Theorem}[section]
  
  \newtheorem{lem}[theorem]{Lemma}

\begin{document}

\title{Fuzzy Line Bundles, the Chern Character and 
Topological Charges over the Fuzzy Sphere}

\author{Harald Grosse$^{a}$, Christian W.~Rupp$^{a,b}$ \\
and Alexander Strohmaier$^{b}$ }

\date{\small
$^a$Institut f\"ur Theoretische Physik, Universit\"at Wien \\
Boltzmanng. 5, A--1090 Wien, Austria \\
$^b$Institut f\"ur Theoretische Physik, Universit\"at Leipzig \\
Augustusplatz 10--11, D--04109 Leipzig, Germany}

\maketitle

\begin{abstract}

Using the theory of quantized equivariant vector bundles over compact coadjoint
orbits we determine the Chern characters of all noncommutative line bundles
over the fuzzy sphere with regard to its derivation-based differential
calculus. The associated Chern numbers (topological charges) arise to be
noninteger, in the commutative limit the well-known integer Chern numbers of
the complex line bundles over the two-sphere are recovered.
\end{abstract}

{\noindent\footnotesize\textit{MSC:} 81R60; 55R10
\\
\textit{Keywords:} Fuzzy Line Bundles; Chern Character; Topological Charges; Fuzzy Sphere}

\section{Introduction and Overview}

Classical gauge field theories exhibit interesting features connected with the
geometry and topology of nontrivial fibre bundles (over space or space--time).
Examples are mono\-pole and in\-stan\-ton solutions.

The Serre--Swan theorem \cite{swan} (cf.~\cite{varilly01} and \cite{connes94})
leads to a complete equivalence between the category of continuous vector
bundles over a compact manifold $M$ and the finitely generated projective modules
over the unital commutative $C^{\star}$--algebra $C(M)$ of continuous
functions over $M$. This circumstance can be generalized to the smooth case
\cite{connes94} and even holds for non compact manifolds. Moreover, the geometry
of $M$ is encoded in $C(M)$.

In noncommutative geometry one proceeds by {\it defining} vector bundles as
finitely generated  projective left, right or bimodules over some
algebra, which is thought of the algebra of functions over some
noncommutative manifold. Accordingly, the {\it geometrical nontriviality} is
purely algebraical  and encoded solely in projective non free modules over the
noncommutative algebra under consideration. 

For the fuzzy sphere this case has first been analyzed by one of us (HG) et.
al. in \cite{grosse96a} (see also \cite{grosse98}), leading to scalar and
spinor field configurations in monopole backgrounds. A different approach
using spectral triples and their Dirac operator-based differential calculi has
been used in \cite{balachandran00} and \cite{balachandran01}. This leads to
integer topological charges similar to \cite{grosse96a}.

We review the definition of the Chern character on projective modules in section
\ref{projmod} and the complex line bundles over the two-sphere in section
\ref{sphere}. In section \ref{fuzzyline} and \ref{results}, we will show that the
Chern character of projective modules over the matrix algebra of the fuzzy sphere
gives, with respect to its free derivation-based differential calculus, rise to
non-integer Chern numbers. In the commutative limit the well-known Chern characters
of complex line bundles over the two-sphere with its integer topological charges
are recovered.


\section{The Chern Character of Projective Modules} \label{projmod}
Let $\mathcal{A}$ be a complex unital not necessarily commutative
$C^{\star}$--algebra and denote $\mathcal{A}\otimes{\cbb}^{n}$ by $\mathcal
{A}^{n}$. Then any projector (selfadjoint idempotent) $p\in M_n(\mathcal{A})$,
where $M_n(\mathcal{A})= \al\otimes M_n(\cbb)$ denotes the $n\!\times\!
n$--matrices with coefficients in $\mathcal{A}$, defines a (finitely generated)
projective left $\mathcal{A}$--module by $E = \mathcal{A}^{n}p$. Elements $\psi$
of  $E$ can be viewed as $\psi\in\mathcal{A}^{n}$ with $\psi p=\psi$. If
$\mathcal{A}$ is further endowed with a differential calculus $({\Omega^*
(\mathcal{A}), \der})$, the Grassmann connection $\nabla: E\rightarrow \Omega^1
(\mathcal{A})\otimes_{\mathcal{A}}E$ of $E$ is defined by $\nabla = p\circ\der$.
It satisfies $\nabla(f\psi)= f\nabla\psi + \der f\otimes_{\mathcal{A}}\psi$ for
all $f\in\mathcal{A}$, $\psi\in E$. After extending $\nabla$ to $\Omega^1
(\mathcal{A})\otimes_{\mathcal{A}}E$ one can define the $\mathcal{A}$-linear map
\beq
\nabla^2:E\rightarrow \Omega^2(\mathcal{A})\otimes_{\mathcal{A}}E,
\eeq
called the curvature of $\nabla$. For more details see \cite{karoubi} and
\cite{landi97}. Evaluating $\nabla^2$ one finds $\nabla^2=p(\der p)(\der p)$,
whereas if we write $(p_{ij})=p\in M_n(\mathcal{A})$, $(\der p)$ is the $n\!
\times\! n$--matrix with coefficients $\der p_{ij}$ and the entries of $p(\der p)
(\der p)$ are $p_{il}\der p_{lk}\wed\der p_{kj}$. The curvature can be viewed as
$\nabla^2\in\Omega^2(\mathcal{A})\otimes_{\mathcal{A}}\End_{\mathcal{A}}(E)$,
where $\End_{\mathcal{A}}(E)$ denotes the right $\mathcal{A}$--module of
endomorphisms of $E$, i.e. $\mathcal{A}$--linear  mappings from $E$ to $E$.
Now define
\beq
\first:=\tr\, p(\der p)(\der p)\in \Omega^2(\mathcal{A}),
\eeq
which is a cocycle, i.e. $\der\first = \tr (\der p)(\der p)(\der p) = 0$. Here
$\tr$ is the trace in $\End_{\mathcal{A}}$. So $\first$ defines a cohomology
class $[\first]\in H^2(\mathcal{A})$. More generally, the Chern character of $E$
with respect to $({\Omega^*(\mathcal{A}), \der})$ is the set of 
\[
\mathrm{Ch}_r(p):=\frac{1}{r!}\tr\, p(\der p)^{2r},\quad r\in\nbb\cup\{0\},
\]
the $\mathrm{Ch}_r(p)$ are called its $r$th compontents. They are also cocycles
and provide equivalence classes in $H^{2r}(\mathcal{A})$. $\mathrm{Ch}_0(p)=\tr\,
p$ simply gives the rank of the module.

\section{Complex Line Bundles over $S^2$} \label{sphere}

One approach to the construction of the complex line bundles over the two-sphere
is the one given in \cite{landi01}, cf. also \cite{landi00}. Starting with the
complex Hopf fibration $U(1)\hookrightarrow SU(2)\simeq S^3\twoheadrightarrow S^2$
and the irreducible representations of $U(1)$ on $\cbb$ labelled by integers
$k\in\zbb$, one defines the space of smooth equivariant functions $C^{\infty}_{(k)}
(S^3,\cbb)\ni\varphi:S^3\rightarrow\cbb$ with $\varphi(x\cdot z)=z^{-k}\varphi(x)$
for $x\in S^3$ and $z\in U(1)$. These are modules over $C^{\infty}_{(0)}(S^3,\cbb)
\simeq C^{\infty}(S^2,\cbb)$, and as such are isomorphic the smooth sections $\Gamma
^{\infty}(S^2, L^k)$ of the associated complex line bundles $L^k:= S^3\times_{k}
\cbb$ over the two-sphere.

By the Serre--Swan theorem this modules are finitely generated and projective and
hence, it is possible to identify $\Gamma^{\infty}(S^2, L^k)$ with $(C^{\infty}
(S^2,\cbb))^n p$, where $p\in M_n(C^{\infty}(S^2,\cbb))$ is a projector. In
\cite{landi01} the projectors $p$ where explicitly constructed with help of the
equivariant functions $C^{\infty}_{(k)}(S^3,\cbb)$. The integer $n\in\nbb$ turned
out to be $|k|+1$ and the first Chern numbers where calculated giving
\[
c_1(p) := \frac{\im}{2\pi}\int_{S^2}\tr\, p(\der p)(\der p) = -k\in\zbb.
\]

Let us shortly mention that $k$ is related to the magnetic charge $Q_m$ of a Dirac
(point) monopole in $\rbb^3$ via
\beq
Q_m = k \frac{\hbar c}{2e},
\eeq
where $\hbar$ is Planck's constant over $2\pi$, $c$ the vacuum speed of light and $e$ the
elementary electrical charge, meaning that $Q_m$ is quantized. 

\section{Fuzzy Line Bundles} \label{fuzzyline}

\subsection{General Remarks} \label{general}

We start with the repetition of well-known facts about the fuzzy sphere and its
free derivation-based differential calculus. Then we use the prescription of
quantizing equivariant vector bundles over coadjoint orbits to obtain the
projectors that define the modules over the matrix algebra of the fuzzy sphere
and its Chern characters.

Denote $\mathrm{SU}(2)$ by $G$, its Lie algebra $\mathrm{su}(2)$ by $\lie$ and let
$\{X_a\}_{a=1,2,3}$ be the generators of the irreducible spin--$N$
representation of $\lie$ acting on the representation space $[N]$ with
$\dim([N]) = 2N+1$.

The algebra of the fuzzy sphere \cite{madore92}, \cite{madore99} is the
noncommutative algebra $\End([N])\simeq\al_N$, the algebra of $(2N+1)
\times(2N+1)$--matrices with complex coefficients. $\al_N$ is generated by
$Y_a = (N(N+1))^{-1/2}X_a$ which satisfy
\beq
[Y_a,Y_b] = \frac{\im\epsilon_{abc}}{\sqrt{N(N+1)}}Y_c \quad\mbox{ and }\quad
\sum_{a=1}^3 (Y_a)^2 = 1.
\eeq

The derivation-based differential calculus (cf. \cite{dubois}) on $\al_N$ is
definded as follows: One chooses the three derivations (``vector fields'') $e_a$,
defined by $e_a(f):=[X_a,f]$. We denote by $\Vef_3(\al_N)$ the linear subspace
of $\Vef(\al_N)$ generated by the $e_a$'s. Here $\Vef(\al_N)$ is the
$\cbb$--vector space of all derivations of $\al_N$. $\al_N$
decomposes into $[0]\oplus[1]\oplus\cdots\oplus [2N]$ as $\lie$-- and $G$--module,
respectively. The derivations $e_a$ satisfy $[e_a,e_b]= \im\epsilon_{abc}e_c$, 
and are the noncommutative analogue of the vector fields $L_a=\im \epsilon_{abc}
x_b\partial /\partial x_c$ on the two-sphere.

The set of $p$-forms $\Omega^p_{(N)}$ over $\al_N$ is the free $\al_N$--module
\beq
\Omega^p_{(N)}=\al_N\otimes(\Vef_3(\al_N)^*\wed\cdots\wed\Vef_3(\al_N)^*)
\simeq\al_N\otimes(\lie_{\cbb}^*\wed\cdots\wed\lie_{\cbb}^*),
\eeq
where $\lie_{\cbb}\simeq\mathrm{sl}(2,\cbb)$. Note that $\Omega^p_{(N)}
=0$ for $p>3$. The exterior derivative $\der:\al_N\rightarrow\Omega^1_{(N)}$ is defined
by $\der \phi(u) = u(\phi)$ for all $\phi\in\al_N$ and $u\in\Vef_3(\al_N)$. It extends to
$\Omega^*_{(N)}=\bigoplus_p\Omega^p_{(N)}$ 
by linearity and the graded Leibniz rule. There is a dis\-tin\-guished
one-form $\Theta$ defined by $\Theta(e_a) = -X_a$. $\Theta$ is the analogue of the
Maurer--Cartan form satisfying $\der\Theta+\Theta^2=0$. The exterior derivative of
a zero form $\phi\in\Omega^0_{(N)}=\al_N$ can with help of $\Theta$ be written as
$\der\phi = -[\Theta,\phi]$. One can choose a basis $\Theta_a$ in $\Omega^1_{(N)}$
completely determined by $\Theta_a(e_b)=\delta_{ab}\one$. Then $\Theta=-X_a\Theta_a$
and $\der\phi = e_a(\phi)\Theta_a$.

It follows from the procedure given in \cite{hawkins99} (cf. also \cite{hawkins00})
that the quantization of equivariant vector bundles over coadjoint orbits is
achieved by means of the orthogonal projection
\[
p\in [N]\otimes[N]^*\otimes[\nu]\otimes[\nu]^*\cong \al_N\otimes\End[\nu],
\]
which projects onto the unique irreducible subrepresentation of $[N]\otimes[\nu]$
with highest (lowest) spin, i.e. onto $[N\pm\nu]$. Here $[\nu]$ is the representation
space of the irreducible spin--$\nu$ representation of $\lie$. In the case
$[N-\nu]$ we have to assume that $N>\nu$. So the fuzzy line
bundles obtained in this way are of the form $\mathbf{L}^{\pm 2\nu}:=
(\al_N\otimes[\nu])p$, they
are isomorphic to the $(2N+1)\times(2(N\pm\nu) +1)$--matrices. Here $p$
acts from the right providing left $\al_N$--modules. The modules $\mathbf{L}^k$ 
approximate in the commutative limit $N\to\infty$ the module of sections of $L^k$. 

Let $\pi:G\rightarrow \End([N\pm\nu])$ be the ir\-re\-ducible representation of
$G$ with spin $N\pm\nu$ and $|h\rangle$ its highest weight vector. $|h\rangle$
is thought of being embedded in $[N]\otimes[\nu]$ as $|h\rangle\oplus 0
\oplus\cdots\oplus 0$. Denote by $\mu$ the normalized Haar
measure on $G$.

\begin{lem} \label{lemmap}
The projector $p:[N]\otimes[\nu]\rightarrow [N\pm\nu]$ defined above is given by
\beq \label{pone}
p=\left(2(N\pm\nu)+1\right)\int_G \pi(g)|h\rangle\langle h|\pi(g)^{-1}d\mu(g).
\eeq
\end{lem}

\begin{proof}
Denote by $p_1$ the right hand side of equation (\ref{pone}). Then $p_1$ sends
every vector in $[N\pm\nu]^{\bot}$ to zero. Now the invariance of $\mu$ implies
that $\pi(g)p_1\pi(g)^{-1} = p_1\:\forall g\in G$, so by the Schur Lemma $p_1$
is proportional to the identity on $[N\pm\nu]$. Since $\tr\: p_1 = 2(N\pm\nu) + 1$,
$p_1$ restricted to $[N\pm\nu]$ is the identity and $p_1^2=p_1$. Accordingly, $p=p_1$.
\end{proof}

\subsection{Explicit Calculations}

For the sake of simplicity we identify in this section $\Vef_3(\al_N)$ with
$\lie_{\cbb}$. It is now our aim to calculate the first component of the Chern
character determined by $p$, i.e. $\first= \tr_2(p\,\der p\,\der p)\in \al_N\otimes
(\lie_{\cbb}^*\wed\lie_{\cbb}^*)$, where `$\der$' acts only on the $\al_N$ part of
$p\in \al_N\otimes \End([\nu])$ and $\tr_2$ is the trace in $\End([\nu])$. 

\begin{lem} \label{F}
$\first = f\epsilon_{abc}X_c\Theta_a\wed\Theta_b$ with $f\in\cbb\,\one$.
\end{lem}

\begin{proof}
Let $\Adj$ be the adjoint representation of $G$ on $\lie_{\cbb}\ni u\mapsto \Adj_g 
u = g u g^{-1}$ and $\Theta$ as defined in section \ref{general}. Then
$\Theta\otimes\one\in\Omega^1_{(N)}\otimes\End([\nu])$ transforms as
\begin{eqnarray*}
\Theta(\Adj_g u)\otimes\one & =
&(\pi_1(g)\otimes\one)(\Theta(u)\otimes\one)(\pi_1^{-1}(g)\otimes\one) \\
& = & \bar{\pi}(g)(\Theta(u)\otimes\one)\bar{\pi}^{-1}(g),
\end{eqnarray*}
where $\bar{\pi} = \pi_1\otimes\pi_2$ is acting on $[N]\otimes[\nu]$ and
$u\in\lie_{\cbb}$, i.e. $\Theta\otimes\one$ is invariant under the action of $G$ on
$\Omega^1_{(N)}\otimes\End([\nu])$. This implies for $\der p =
-[\Theta\otimes\one,p]$ that $\der p(\Adj_g u) = \bar{\pi}(g)\der p(u)
\bar{\pi}^{-1}(g)$
and finally for the first component of the Chern character
\[
\first(u,v) = \pi_1(g)\first(\Adj_{g^{-1}}u,\Adj_{g^{-1}}v)\pi_1^{-1}(g)
\]
for all $u,v\in\lie_{\cbb}$. So $\first$ is an invariant element of
$\al_N\otimes(\lie_{\cbb}^*\wed\lie_{\cbb}^*)$. Reducing this latter space as a
$G$--module shows that $[0]$ appears only once, as $\lie_{\cbb}^*\wed\lie_{\cbb}^*
\simeq [1]$. Consequently, the subspace of invariant two-forms is one dimensional and,
since $\epsilon_{abc}X_c\Theta_a\wed\Theta_b$ is invariant, $\first$ can be
written as claimed.
\end{proof}
Note that $f\epsilon_{abc}X_c\Theta_a\wed\Theta_b$ can also be written as 
$iq/4\,\epsilon_{abc}Y_a\der Y_b\wed\der Y_c$, where $q$ and $f$ are related by
\[
q = \frac{4}{\im}\frac{(N(N+1))^{3/2}}{1/2 - N(N+1)}f,
\]
as can be seen by expanding $\der Y_a=[X_b,Y_a]\Theta_b$. It will turn out later that
$q$ can be naturally interpreted as Chern numbers.

What is left to do is to determine $f$, depending on $N$ and $\nu$.
For this note first that
\[
\tr_2 (p\,\der p(e_a)\der p(e_b)) = f\epsilon_{abd}X_d.
\]
Now multiply this equation with $\epsilon_{abc}X_c$ and take the trace also in
$\al_N$ to get
\beq \label{f}
f = \frac{\epsilon_{abc}\tr (p\,\der p(e_a)\der p(e_b)X_c)}{2N(N+1)(2N+1)},
\eeq
where $\tr$ denotes the trace in $\End([N])\otimes\End([\nu])$. This expression
can be further simplified by the following lemma:

\begin{lem} \label{expect}
\[
\epsilon_{abc}\tr (p\,\der p(e_a)\der p(e_b)X_c) = (2(N\pm\nu)+1)
\epsilon_{abc}\langle h|[X_a,p][X_b,p]X_c|h\rangle.\]
\end{lem}

\begin{proof}
With $\bar{\pi}$ as above the left hand side is by lemma \ref{lemmap}
\beq \label{int}
(2(N\pm\nu) +1)\int_G\epsilon_{abc}\langle
h|\bar{\pi}^{-1}(g)[X_a,p][X_b,p]X_c\bar{\pi}(g)|h\rangle d\mu(g).
\eeq
Now $\epsilon_{abc}[X_a,p][X_b,p]X_c$ are the components of
the equivariant multiliear map $-[\Theta\otimes\one,p][\Theta\otimes\one,p]
(\Theta\otimes\one)$ from $\lie_{\cbb}\wed\lie_{\cbb}\wed\lie_{\cbb}\simeq [0]$
to $\al_N\otimes\End([\nu])$ which is constant. This implies the assertion.
\end{proof}

Let us denote the expectation value $\epsilon_{abc}\langle h|\cdot|h\rangle$
appearing in lemma \ref{expect} by $B$. Expanding the commutators it is
straightforward to see that $B=C+iD$ with
\beq \label{cd}
C = \epsilon_{abc}\langle h|X_a p X_b p X_c|h\rangle \quad\mbox{and}\quad 
D = -\langle h| X_a p X_a|h\rangle.
\eeq
Consider now (cf. \cite{gitman93}) the space of homogenous polynomials $\poly{n}$ of
two complex variables $z_1$ and $z_2$ of fixed degree $n\in \nbb$. We define the
following ``creation'' and ``annihilation'' operators $a_i^\dagger = z_i$ and $a_i = 
\partial / \partial z_i$, satisfying $[a_i,a_j^\dagger] = \delta_{ij}$, which give an
irreducible representation of $\lie$ by
\[
X_1 = \frac{1}{2} (a_1^\dagger a_2 + a_2^\dagger a_1), \ 
X_2 = -\frac{\im}{2} (a_1^\dagger a_2 - a_2^\dagger a_1) \ \mbox{and}\ 
X_3 = \frac{1}{2} (a_1^\dagger a_1 - a_2^\dagger a_2)
\]
with spin $n/2$ on $\poly{n}$. To compute $C$ and $D$ of equation (\ref{cd}) we
realize the $[N]\otimes[\nu]$ representation on $\poly{n}\otimes\poly{l}$ with
$2N=n$ and $2\nu=l$, respectively. An orthonormal basis of $\poly{n}$ is given by
\beq
|\psi_k\rangle = {n \choose k}^{1/2}z_1^k z_2^{n-k}\quad\mbox{ with }\quad
\langle \psi_k|\psi_{k'} \rangle = \delta_{kk'}.
\eeq

First we analyze the case where $p$ projects onto $[N+\nu]$. Then the highest
weight vector $|h\rangle\in\poly{n}\otimes\poly{l}$ is
given by $|h\rangle=z_1^n\otimes z_1^l$, i.e. $(X_3\otimes\one + \one\otimes
X_3)|h\rangle = \oh (n+l)|h\rangle$ and $(X_+\otimes\one + \one\otimes
X_+)|h\rangle=0$, with $||h||=1$.
Define $|w\rangle := X_1|h\rangle$, then $X_2|h\rangle = \im |w\rangle$. Since
$X_3|h\rangle = \oh n|h\rangle$ a lenghty but straightforward calculation yields
\beq \label{B}
B = 2\im(n-1)\langle w|p|w\rangle - 2\im\langle w|pX_3p|w\rangle - \frac{\im n^2}{4}.
\eeq
Now because $|w\rangle$ and $|v\rangle := (X_-\otimes\one + \one\otimes
X_-)|h\rangle$ have the same eigenvalue $\oh(n+l-2)$ of $X_3\otimes\one +
\one\otimes X_3$, we know that $p|w\rangle = \lambda |v\rangle$. Accordingly,
$\lambda$ can be evaluated
\[
\lambda = \frac{\langle v|w\rangle}{\langle v|v\rangle} =
\frac{1}{2}\frac{n}{n+l}.
\]
This gives
\beq \label{wpw}
\langle w|p|w\rangle = \frac{1}{4}\frac{n^2}{n+l}\quad\mbox{ and }\quad
\langle w|pX_3p|w\rangle = \frac{1}{8}\frac{n^2}{(n+l)^2}(n(n-2) + nl).
\eeq
Inserting (\ref{wpw}) into (\ref{B}) this gives finally for $f$ in equation
(\ref{f}) expressed in terms of $N$ and $\nu$
\beq
f = -\im N \nu \frac{(N+\nu+1)(N+\nu+1/2)}{(N+\nu)^2(2N+1)(N+1)}.
\eeq

The case where $p$ projects onto $[N-\nu]$ is more involved, since we first have
to determine the highest weight vector. The ansatz
\[
|h\rangle = \sum_{k=0}^l a_k z_2^k z_1^{n-k}\otimes z_1^k z_2^{l-k}
\]
leads through $(X_+\otimes\one + \one\otimes X_+)|h\rangle = 0$ to the recursion
relation $(k-l)a_k = (k+1)a_{k+1}$ which is solved by
\beq
a_k = (-1)^ka_0 {l \choose k}.
\eeq
The remaining $a_0$ is determined by the normalization condition $||h||=1$ and
gives for $a_k$
\beq
a_k = (-1)^k\sqrt{\frac{n-l+1}{n+1}}{l \choose k}\quad\mbox{ for } k=0,\ldots,l,
\eeq
where we have used the formula
\beq
\sum_{k=0}^l {l \choose k}{n\choose k}^{-1} = \frac{n+1}{n-l+1}.
\eeq 
Proceeding analogously we define $|w\rangle = X_1|h\rangle$, but now
$|w\rangle = |w_+\rangle + |w_-\rangle$ with
\[
(X_3\otimes\one + \one\otimes X_3)|w_{\pm}\rangle = \left( \frac{n-l}{2} \pm 1
\right)|w_{\pm}\rangle.
\]
From this it follows that $pX_1|h\rangle = p|w_-\rangle = \lambda|v\rangle$ and 
$pX_2|h\rangle = \im  p|w_-\rangle = \im\lambda|v\rangle$, here
$|v\rangle: = (X_-\otimes\one + \one\otimes X_-)|h\rangle$. Using
\[
\sum_{k=0}^l {l\choose k}{n \choose k\!+\!1}^{\!-1} = \frac{n+1}{(n-l)(n+1-l)}
\quad\mbox{ and }
\]
\beq \label{binom}
\sum_{k=0}^l(n-k){l \choose k}{n\choose k\!+\!1}^{\!-1} =
\frac{(n+1)(n+2)}{(l-n-1)(l-n-2)}
\eeq
one finds for the proportionality factor $\lambda$
\beq
\lambda = \frac{\langle v|w_-\rangle}{\langle v|v\rangle} =
\frac{1}{2}\frac{n+2}{n-l+2}.
\eeq
Applying $X_3$ to $|h\rangle$ yields $\oh n|h\rangle - |K\rangle$ with
\[
|K\rangle = \sum_{k=1}^l a_k k z_2^k z_1^{n-k}\otimes z_1^k z_2^{l-k}.
\]
Since $(X_3\otimes\one + \one\otimes X_3)|K\rangle = \oh(n-l)|K\rangle$ it
follows that $pX_3|h\rangle = (\oh n - \mu)|h\rangle$, where $\mu$ is given by
\[
\mu = \langle h|K\rangle = \frac{l}{n+2-l},
\]
for which a formula similar to (\ref{binom}) has been used. Now we have enough
ingredients to write $B$ as
\beq
B = 2\im\lambda^2 \Big( (n - 2\mu -1)\langle v|v\rangle - \langle
v|X_3|v\rangle\Big) - \im\left( \frac{n}{2} - \mu \right)^2.
\eeq
It is left is to determine $\langle v|X_3|v\rangle$. One finds
\begin{eqnarray*}
\langle v|X_3|v\rangle & = &\frac{n+1-l}{n+1}\frac{(n-l)^2}{2}
\sum_{k=0}^l{l\choose k}{n\choose k\!+\!1}^{\!-1}(n-2k-2) \\
& = & \frac{(n+2)(n-l-2)(n-l)}{2(n-l+2)}.
\end{eqnarray*}
We collect the results to obtain $f$ in terms of $N$ and $\nu$
\beq
f = \frac{\im\nu(N+1)(N-\nu)(2N-2\nu+1)}{2N(2N+1)(N-\nu+1)^2}.
\eeq

\section{Results and Commutative Limit} \label{results}
Summarizing the calculations of the previous chapter we get for the Chern
character $\first$ of the modules $\mathbf{L}^k$, with $k=\pm 2\nu$,
the formula
\beq
\first = \frac{\im q\epsilon_{abc}}{4}Y_a\der Y_b\wed\der Y_c,
\eeq
where
\beq
q = \frac{4}{\im}\frac{(N(N+1))^{3/2}}{1/2 - N(N+1)}f,
\eeq
and
\begin{eqnarray}
f & = & \frac{\im\nu(N+1)(N-\nu)(2N-2\nu+1)}{2N(2N+1)(N-\nu+1)^2} \quad
\mbox{for } k=2\nu>0, \\
f & =  & -\im N \nu \frac{(N+\nu+1)(N+\nu+1/2)}{(N+\nu)^2(2N+1)(N+1)}
\quad \mbox{for } k=-2\nu<0. 
\end{eqnarray}

What is needed to obtain the associated Chern numbers is a certain notion of
integration over two-forms. Thus for $\phi\in {\mathcal A}_N$ and
\[
\omega:=\frac{\epsilon_{abc}Y_a}{8\pi} \der Y_b\wed \der Y_c \in \Omega^2({\mathcal A}_N)
\]
we define the integral by
\beq
\int^{\star} \phi\,\omega := \tr_{\!N} (\phi),\:\:\mbox{ where } 
\tr_{\!N}(\cdot)=\frac{1}{2N+1}\tr(\cdot),
\eeq
with $\int^{\star}\omega =1$. The two-form $\omega$ is the noncommutative volume 
form, which in the commutative limit converges to the normalized volume form on 
$S^2$. Consequently, the first Chern numbers of the fuzzy line bundles determined by
$p$ are given by
\beq
c_1(p) := \frac{\im}{2\pi}\int^{\star}\first = -q.
\eeq
In the continuum limit we find for the topological charges
\beq
k=\lim_{N\to \infty} c_1(p) = \mp 2\nu\in\zbb,
\eeq
where the minus sign corresponds to the projection onto $[N+\nu]$ and the plus
sign to $[N-\nu]$. Some topological charges $q$ and their commutative limits
$k$ are shown in Figure 1. 

\section{Conclusions}
We constructed projective modules over the matrix algebra $\al_N$ of the fuzzy
sphere using the prescription of quantizing equivariant vector bundles given in
\cite{hawkins99}, leading to fuzzy line bundles. With respect to the free
derivation-based differential calculus $(\Omega^*_{(N)},\der)$ on $\al_N$ we calculated
the Chern
character $\first\in\Omega^2_{(N)}$. Since $\first$ was seen to be $\mathrm{SU}
(2)$--invariant, i.e. an $\mathrm{SU}(2)$--equivariant mapping from $\mathrm{sl}
(2,\cbb)\wed\,\mathrm{sl}(2,\cbb)$ to $\al_N$, it was unique up to a factor. The
determination of this factor $f\in\cbb$ was achieved and with help of a certain
notion of integration the Chern numbers $q$ associated with $\first$ were calculated.
These turned out to be non-integer, becoming integers in the commutative limit
$N\rightarrow\infty$.
\section{Acknowledgements}
This work was partly supported by the ``Deutsche Forschungsgemeinschaft'' within
the scope of the ``Graduiertenkolleg Quantenfeldtheorie'' of the University of Leipzig.
CWR and AS want to thank J. Dietel and S. Kolb for pleasant discussions.


\begin{figure}
\centerline{
\psfrag{c}{$q$}
\psfrag{f}{$1/N$}
\includegraphics[width=10cm]{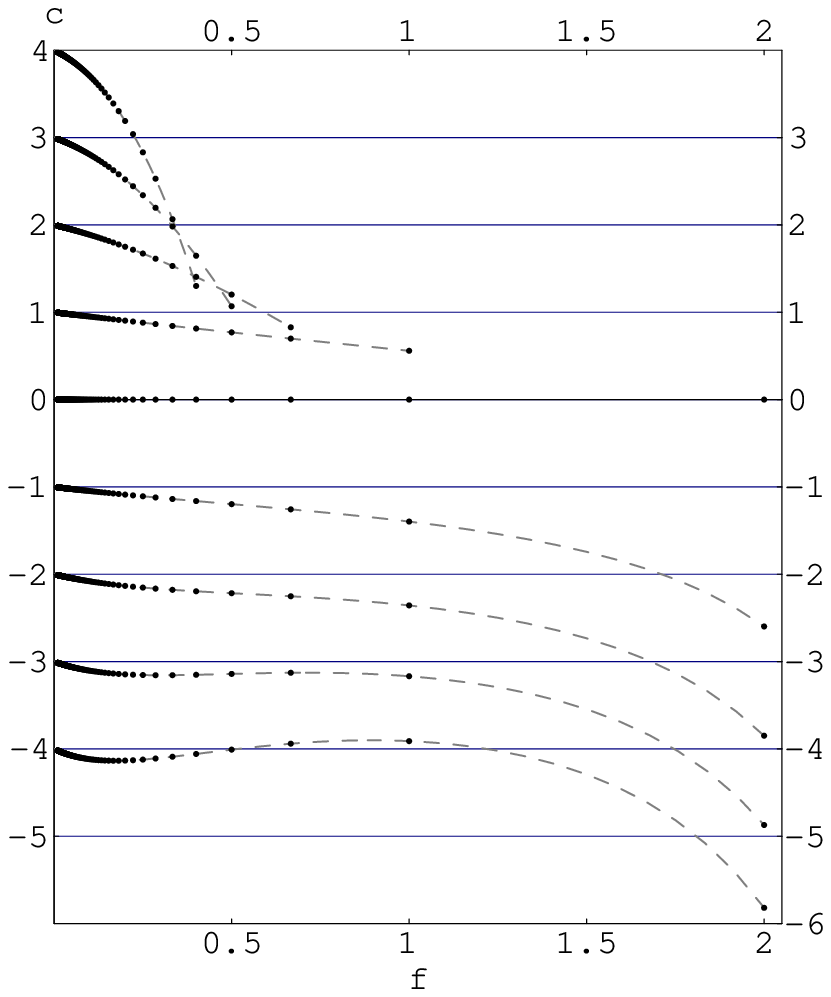}}
\vspace{0.1cm}
\caption{Topological charges $q$ with commutative limit $k$ between $-4$ and $4$
as function of the fuzzyness $1/N$. These can be viewed as the (fuzzy)
magnetic charges of a Dirac monopole living on the fuzzy sphere. Dashed lines connect
charges of constant $\nu$.}
\label{charge}
\end{figure}
\end{document}